\documentstyle[11pt,aas2pp4,epsfig]{article}
\begin{document}

\title{The Long Term Stability of Oscillations During Thermonuclear
X-ray Bursts: Constraining the Binary X-ray Mass Function}
\author{Tod E. Strohmayer,  William Zhang, Jean
H. Swank}
\affil{Laboratory for High Energy Astrophysics, Goddard Space Flight Center,
Greenbelt, MD 20771}
\author{Iosif Lapidus}
\affil{Univeristy of Sussex, Astronomy Centre, Physics and Astronomy Subject
Group, CPES, Falmer, Brighton BN1 9QH, England}
\authoraddr{Laboratory for High Energy Astrophysics, Mail Code 662, NASA/GSFC
Greenbelt, MD 20771}
\vskip 10pt

\centerline{Accepted for publication in the Astrophysical Journal Letters}

\begin{abstract}

We report on the long term stability of the millisecond oscillations
observed with the Rossi X-ray Timing Explorer (RXTE) during thermonuclear 
X-ray bursts from the low mass X-ray binaries (LMXB) 4U 1728-34 and 4U 1636-53.
We show that bursts from 4U 1728-34 spanning more than 1.5 years have observed
asymptotic oscillation periods which are within 0.2 $\mu$sec of each
other, well within the magnitude which could be produced by the orbital motion
of the neutron star in a canonical LMXB. This stability implies a timescale to
change the oscillation period of $> 23,000$ years in this system, and suggests a
highly stable process, such as stellar rotation, as the mechanism producing the
oscillations. For 4U1636-53, which has an orbital period of 3.8 hours, we show
that offsets in the asymptotic oscillation periods from three different bursts
can be consistently interpreted as due to orbital velocity of the neutron star
with $v \sin i /c \approx 4.25 \times 10^{-4}$. An updated optical ephemeris for
the epoch of maximum light from V801 Arae would provide a strong test of this
interpretation. We discuss the constraints on the X-ray mass function which can
in principle be derived using this technique.

\end{abstract}

\vfill\eject

\section{Introduction}
Millisecond oscillations in the X-ray brightness during thermonuclear bursts 
have now been observed from 6 low mass X-ray binaries (LMXB) with the Rossi
X-ray Timing Explorer (RXTE) (see \markcite{Stroh96}Strohmayer {\it et al.}
1996; \markcite{SMB}Smith, Morgan \& Bradt 1997; \markcite{Stroh97}Strohmayer
{\it et al.} 1997; \markcite{Z96}Zhang {\it et al.} 1996; \markcite{Z98}
Zhang {\it et al.} 1997; and \markcite{SZS}Strohmayer, Zhang \& Swank 1997).
The presence of large amplitude oscillations near burst onset, combined with
spectral evidence for localized thermonuclear burning suggests that the 
oscillations are caused by rotational modulation of thermonuclear 
inhomogeneities (see \markcite{SZW}Strohmayer, Zhang \& Swank 1997).

The accretion-induced rate of change of the neutron star spin frequency in a 
LMXB is approximately 
\begin{equation}
d\nu / dt \approx 1.8 \times 10^{-6}  \; \frac{\dot m_{17} (M_{x}
r_{acc})^{1/2}} {2\pi I_{45}} {\rm Hz \; yr}^{-1} \;,
\end{equation}
where $\dot m_{17}$, $M_{x}$, $r_{acc}$ and $I_{45}$ are the mass accretion
rate in units of $10^{17}$ g s$^{-1}$, the neutron star mass in solar units, 
the characteristic radius of the inner accretion disk in km, and the stellar
moment of inertia in units of $10^{45}$ g cm$^2$, respectively. If the
millisecond oscillations observed in the X-ray brightness from thermonuclear
X-ray bursts with the Rossi X-ray Timing Explorer (RXTE) are produced by
rotational modulation of the burst flux, then the Doppler corrected
frequencies should be stable at the better than $\Delta\nu = 0.001$ Hz level
over a hundred years or so. The Doppler shift due to orbital motion of the
binary can produce a frequency shift of magnitude 
\begin{equation}
\Delta\nu / \nu = v \sin i /c = 2.05 \times 10^{-3} \;\frac{M_c \sin i}
{P_{hr}^{1/3} (M_x + M_c)^{2/3}} \; ,
\end{equation}
where $M_x$, $M_c$, $P_{hr}$, $v$ and $i$ are the neutron star mass,
the companion mass (both in solar units), the orbital period in hours, the 
magnitude of the neutron star orbital velocity, and the
system inclination angle, respectively. For canonical LMXB system parameters 
this doppler shift easily dominates over any possible accretion-induced spin
change on orbital to several year timescales. Thus, the level of observed
stability in oscillation periods from burst to burst provides a 
method to further test the rotational modulation hypothesis. For example, 
if oscillation period shifts larger than can plausibly be produced 
via orbital motion are observed this would tend to cast doubt on the spin
modulation interpretation. On the other hand, if the burst oscillation
frequencies remain stable over long timescales, revealing a signature of binary
motion, then it will both support the rotational interpretation and become
possible to use the observed frequency shifts to constrain the neutron star
binary mass function in systems which have observed burst oscillations.

In this Letter we investigate the long term period stability of burst
oscillations in two LMXB sources, 4U 1728-34 and 4U 1636-53. We show that in 
4U 1728-34 bursts separated in time by about 1.6 years have shortest observed
oscillation periods, what we refer to as asymptotic periods, within
0.2  $\mu$sec of each other, well within the range of shifts which could 
result from the system's binary motion. For 4U 1636-53, which has a known
orbital period of 3.8 hours, our time baseline is shorter,
however, the orbital period allows us to show that the oscillation 
period shifts observed from three different bursts can be consistently
interpreted in terms of those produced by binary orbital motion with reasonable
values for the component masses and system inclination. If the relative orbital
phases when the bursts ocurred can be converted to absolute phases, for example,
with an updated optical ephemeris of the epoch of maximum light, then this 
would provide a test of the doppler shift interpretation, and if confirmed would
enable constraints on $v \sin i / c$ and thus the X-ray mass function
to be derived.

\section{Long Term Frequency Stability in 4U1728-34}

We had observations of 4U1728-34 with RXTE in February, 1996 and again in 
September, 1997 (see \markcite{Stroh96}Strohmayer {\it et al.} 1996 for a 
summary of the February, 1996 observations). Using data from these two 
observations we can compare the oscillation periods during bursts 
over a span of about 1.6 years. For all the burst data reported here we
had either 125 $\mu$sec (1/8192 s) resolution binned data or event mode
data with the same temporal resolution. 

The oscillations at 2.75 ms (363 Hz) observed during bursts from 4U 1728-34 are
not strictly coherent (see \markcite{Stroh96}Strohmayer {\it et al.} 1996).
In some bursts the period is observed to evolve 
from a high of about 2.762 ms near burst onset to about 2.747 ms during burst
decay. Due to the episodic nature of the oscillations not all bursts show
detectable oscillations over this entire range. \markcite{Stroh97}Strohmayer
{\it et al.} (1997) have argued that this frequency evolution is caused by the
increase in the scale height of the thermonuclear burning layer on the neutron
star surface and subsequent conservation of angular momentum of the
thermonuclear shell. In many bursts which show oscillations the oscillation
period appears to reach a nearly coherent, asymptotic limit as the burst decays
away. In the context of the spin modulation hypothesis this limit represents the
actual spin period of the bulk of the neutron star.

We selected for detailed comparison a pair of bursts from the February, 1996
observations (bursts 4 and 5, respectively, from \markcite{SZS}Strohmayer, 
Zhang \& Swank 1997) and one from the September, 1997 data. Here we refer to
these bursts as bursts 1, 2 and 3, in time order. We selected these
bursts because they showed significant oscillations over the longest time
intervals during the bursts and the oscillation period during the burst 
decay reached a stable, coherent limit. 
In Figure 1 we compare the dynamic power spectra of the bursts detected on 
Feb. 16, 1996 at 06:51:07 UTC and Sep. 22, 1997 at 06:42:51 UTC (bursts 2 and 3
in Table 1). We only show two of the three bursts since the pair of bursts
observed on Feb. 16, 1996 were nearly identical in their oscillation properties.
The figure shows contours of constant power spectral amplitude and they have
been shifted in time for clarity. The dynamic power spectra were computed 
from 2 s intervals with a new interval starting every 1/8 s. 
The leftmost contours are for the Feb., 1996 burst. The frequency evolution from
low to high is clearly evident in both bursts, and the range of observed
frequencies and the highest observed frequency are very similar. Note that the 
oscillation frequencies in both bursts reach a stable upper limit, what we 
will call the asymptotic frequency or period.

Since the rotational modulation hypothesis suggests that the shortest observed
period is the underlying stellar spin period we performed an epoch folding
period search analysis using only the portions of all three bursts after which
the frequency has stabilized to see how closely these asymptotic frequencies
agree. Figure 2 shows the resulting $\chi^2$ plots from the folding analysis as
a function of barycentric period. To estimate the oscillation periods and 
uncertainties from the epoch folding we computed the centroids $P_{cen}$ and
standard deviations $\sigma_P$ of each $\chi^2$ peak using the relations
\begin{equation}
P_{cen} = \frac{\sum_i \chi^2_i P_i}{\sum_i \chi^2_i} \;\;\; \sigma^2_P = 
\frac{\sum_i \chi^2_i (P_i - P_{cen})^2}{\sum_i \chi^2_i} \; , 
\end{equation}
where $i$ runs over each bursts $\chi^2$ peak from the epoch folding analysis.
Table 1 summarizes the derived asymptotic periods and uncertainties for each
burst. As can be seen from the inferred periods and uncertainties there is no
significant evidence that the observed asymptotic periods from the three bursts
are different. Using the measured centroids as the best estimator of the periods
for each burst, the implied period difference over the 1.6 yr timespan is about
0.19 $\mu$sec. In terms of a timescale to change the period
this corresponds to $\tau > P / \dot P = P \Delta T / \Delta P = 2.3 \times 
10^4$ yr, and implies a limit on any orbital doppler shift $\Delta P / P = 
v_{orb}\sin i / c < 6.9 \times 10^{-5}$, well within the shift which could be 
produced by orbital motion of the neutron star in a typical LMXB. 

\section{Burst Oscillation Frequencies in 4U 1636-53}

X-ray brightness oscillations during X-ray bursts at 1.72 ms (581 Hz) were
discovered in 4U 1636-53 by \markcite{Zhang96}Zhang {\it et al.} (1996). 
The 3.8 hr orbital period of 4U 1636-53 is known from the observed optical
periodicity of the optical companion V801 Arae (see \markcite{vP90}van Paradijs
{\it et al.} 1990; \markcite{SM88}Smale \& Mukai 1988; and
\markcite{PvL81}Pedersen, van Paradijs \& Lewin 1981). Since the orbital period
is known the {\it relative} phases at which bursts occurred can be determined.
One can then compare the observed oscillation periods from different
bursts and determine if any observed changes can be consistent with an
orbitally induced doppler shift. In particular, if at least three bursts are
available with measured oscillation periods then one can try to solve the
following set of equations;
\begin{equation}
P_{t_j} = P_0 - \Delta P \cos (\phi_{t_j} + \phi_0 ) \; .
\end{equation}
Here, $P_{t_j}$ and $\phi_{t_j}$ are the observed asymptotic oscillation
periods and relative orbital phases, respectively, for bursts which occurred
at $t_j$, and $P_0$, $\Delta P$, and $\phi_0$ are the barycentric oscillation
period when the neutron star transits the line of sight, the magnitude of the
doppler induced period change, and an initial phase offset, respectively.
With at least three different oscillation period measurements during bursts
it may be possible to determine a set of values for the three parameters
$P_0$, $\Delta P$, and $\phi_0$ which are consistent with binary motion. 
The orbital velocity $v$ and system inclination $i$ are related by 
$\Delta P / P_0 = v \sin i / c$. 

For 4U 1636-53 we now have 3 different bursts spanning a time interval of 
slightly more than a day. To determine the asymptotic oscillation periods for
each burst we performed a similar epoch folding analysis on these bursts as 
those from 4U 1728-34 described above. Table 1 summarizes information on the
occurrence times, relative orbital phases (from burst 1) measured at the solar
system barycenter of 4U1636-53 at the time of
occurrence, and the barycentric asymptotic period observed during the decaying
portion of each burst. Figure 3 displays the resulting $\chi^2$ plots from the
epoch folding analysis for each of the three bursts. 

Bursts 1 and 3 occurred approximately half an orbit apart from each other and
note that these bursts had asymptotic periods which are, within the
uncertainties, consistent with each other. Burst 2 occurred roughly midway in
orbital phase between bursts 1 and 3 but had a significantly shorter period 
by about 0.74 $\mu$sec. Thus, a plausible
scenario is that burst 1 occurred near the time of superior conjuction of the
neutron star (inferior conjunction of the secondary). 
At this phase, as well as half an orbit away, the neutron star is
near transit and the orbital component of its velocity along the line of sight
is nearly zero. This can explain the consistent periods measured during bursts
1 and 3. Finally, burst 2 occurred $0.327$ in phase from burst 1 and thus with
a significant fraction of its total orbital velocity
along the line of sight to produce the strong blue shift to shorter period.

This scenario gives $v \sin i /c \approx 4.25 \times 10^{-4}$ to account for
the observed period shift between burst 2 and bursts 1 and 3. Current
understanding of the mechansim underlying the optical modulations at the 
orbital period suggests that the epoch of maximum light be identified with
superior conjunction of the optical secondary (see \markcite{SM88}Smale \&
Mukai 1988). \markcite{vP90}Van Paradijs {\it et al.} (1990) give an ephemeris
for the epoch of maximum light for V801 Arae based on a compilation of
observations from as early as July, 1980 to as late as May, 1988. 
However, the uncertainties
in projecting this ephemeris forward to the epoch of burst 1 are such that it
does not provide a significant test of the orbital doppler interpretation,
but, with an updated optical ephemeris a much more restrictive test can be made.

Since the orbital motion hypothesis cannot be rejected with the present data we
show in figure 4 the contours of constant $\sin i$ in the mass plane for the
appropriate orbital period of $3.8$ hr and the $v \sin i /c$ suggested by the
period shifts in the bursts. For the 3.8 hour orbital period and a main
sequence companion, empirical mass-radius relations would suggest a mass of
about 0.36 $M_{sun}$ for V801 Ara (see \markcite{P84}Patterson 1984; 
\markcite{SM88}Smale \& Mukai 1988). This is a bit lower in mass than that
suggested by the inferred $v \sin i /c$ from the bursts (see Figure 4), 
but given the uncertainties in the period measurements as well as the uncertain
effects of X-ray heating on the secondary, it is still within a reasonable 
range for the observed shifts to be plausibly produced by orbital motion. 
Both the observation of additional bursts as well as an updated optical
ephemeris for V801 Ara can provide a careful test of the orbital hypothesis for
the observed burst oscillation period shifts.  

\section{Discussion}

The episodic nature of the oscillations during bursts does introduce some 
uncertainty into what is the ``highest" observed frequency during a burst. It
is possible that in some bursts the oscillations are not strong enough to
be detected at late times in the burst and therefore the highest frequency may
not be observed in all bursts. Partly this can be mitigated by comparing
bursts which show similar overall oscillation properties, as we have endeavored
to do here, but in order to fully overcome this one simply needs the weight of
evidence from a larger sample of bursts. 
In particular, with a large enough sample to 
cover most of the orbital phase space the signature of orbital doppler shifts
should become fairly transparent, or not, since the magnitude of the frequency
offsets should be limited by the magnitude of the binary orbital velocity, and
an approximately equal number of redshifts and blueshifts should be observed. 

The long term stability of the highest millisecond oscillation frequencies
observed in thermonuclear bursts from 4U1728-34 and 4U1636-53 provides a strong
argument in favor of a higly stable clock, such as stellar rotation, setting 
the observed oscillation frequency. Regardless of the mechanism, any oscillation
period will suffer orbital doppler effects. The limits on the period offsets
from bursts spanning 1.6 years in 4U 1728-34 indicates that the intrinsic
period which sets the asymptotic period during bursts can change on
a timescale no shorter than $\tau = P \Delta T / \Delta P \approx 2.3 \times
10^4$ yrs. This timescale is longer than similar timescales for many known
X-ray pulsars, and is also longer than the characteristic time to change the
thermal state of the neutron star surface ocean (see \markcite{BCUC98}Bildsten,
{\it et al.} 1998). Thus if oscillation modes sensitive to the thermal
state, such as $g$-modes, were the cause of the oscillations, they would not be
expected to be stable over such a long timescale. If analysis of 
additional bursts continues to support this interpretation, then it will 
become possible to use oscillation periods observed in different bursts to
place constraints on the masses of the components in LMXB, thus long pointed
observations that collect many bursts are well justified given that they could
lead to a determination of the mass function for a larger sample of systems.
In addition, constraints on $v \sin i /c$ derived from different bursts 
provides a method to conduct more sensitive searches for the millisecond X-ray
pulsar in the persistent, accretion-driven flux, which should be present at 
some level in most LMXB.

\acknowledgements

We thank Nick White, Glenn Allen and Mike Stark for helpful discussions and 
comments on the manuscript.

\vfill\eject
\onecolumn

\hoffset=-0.4in
\begin{table}
\caption{Asymptotic Oscillation Periods in bursts from 4U1728-34 and 4U1636-53}

\begin{tabular}{ccccc}

Source & $T_{burst}$ (UTC) & $P_{cen}$ (ms) & $\sigma_P$ ($\mu$sec) &
$\phi_{rel} (TBD)$ \cr
\tableline
4U 1728-34 &  &  &  &  \cr
\tableline
burst 1 & 2/16/96 at 06:51:13 & 2.74774 & 0.40 & NA \cr

burst 2 & 2/16/96 at 10:00:49 & 2.74757 & 0.39 & NA \cr

burst 3 & 9/22/97 at 06:42:56 & 2.74755 & 0.39 & NA \cr
\tableline
4U 1636-53 &  &  &  &  \cr
\tableline
burst 1 & 12/28/96 at 22:39:24 & 1.71970 & 0.43 & 0.0 \cr

burst 2 & 12/28/96 at 23:54:03 & 1.71896 & 0.39 & 0.327 \cr

burst 3 & 12/29/96 at 23:26:47 & 1.71981 & 0.38 & 0.517 \cr
\tableline

\end{tabular}
\end{table}

\vfill\eject
\hoffset=0.0in

\twocolumn

\vskip 10pt

\vfill\eject

\begin{figure*}[htb] 
\centerline{\epsfig{file=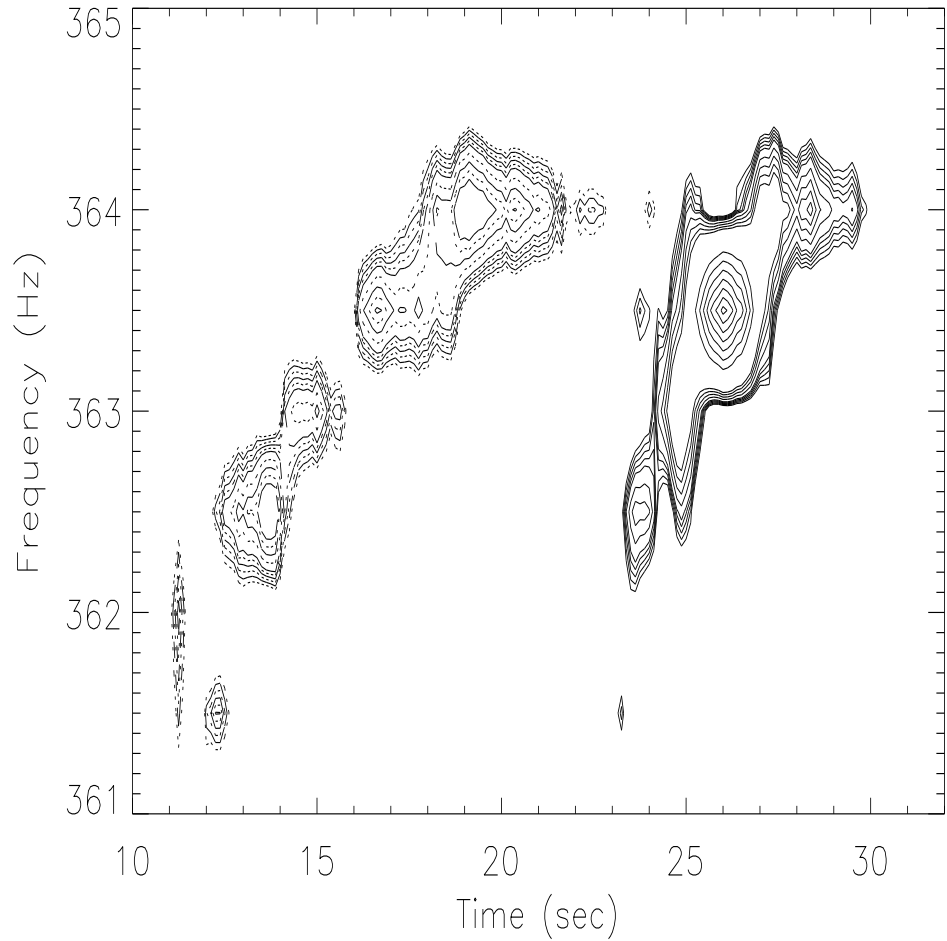,height=5.5in,width=5.5in}}
\vspace{10pt}
\caption{Dynamic power spectra computed from two different bursts
from 4U 1728-34 separated in time by 1.6 yr. Shown are contours of constant
power spectral density. The contours have been offset from each other for
clarity. Note that the range in frequency of the oscillations as well as the
highest observed frequency are very similar. The burst from 2/16/96 at 10:00:49
UTC is on the left, that from 9/22/97 at 06:42:56 UTC in on the right.}
\label{fig1}
\end{figure*}

\vfill\eject

\begin{figure*}[htb] 
\centerline{\epsfig{file=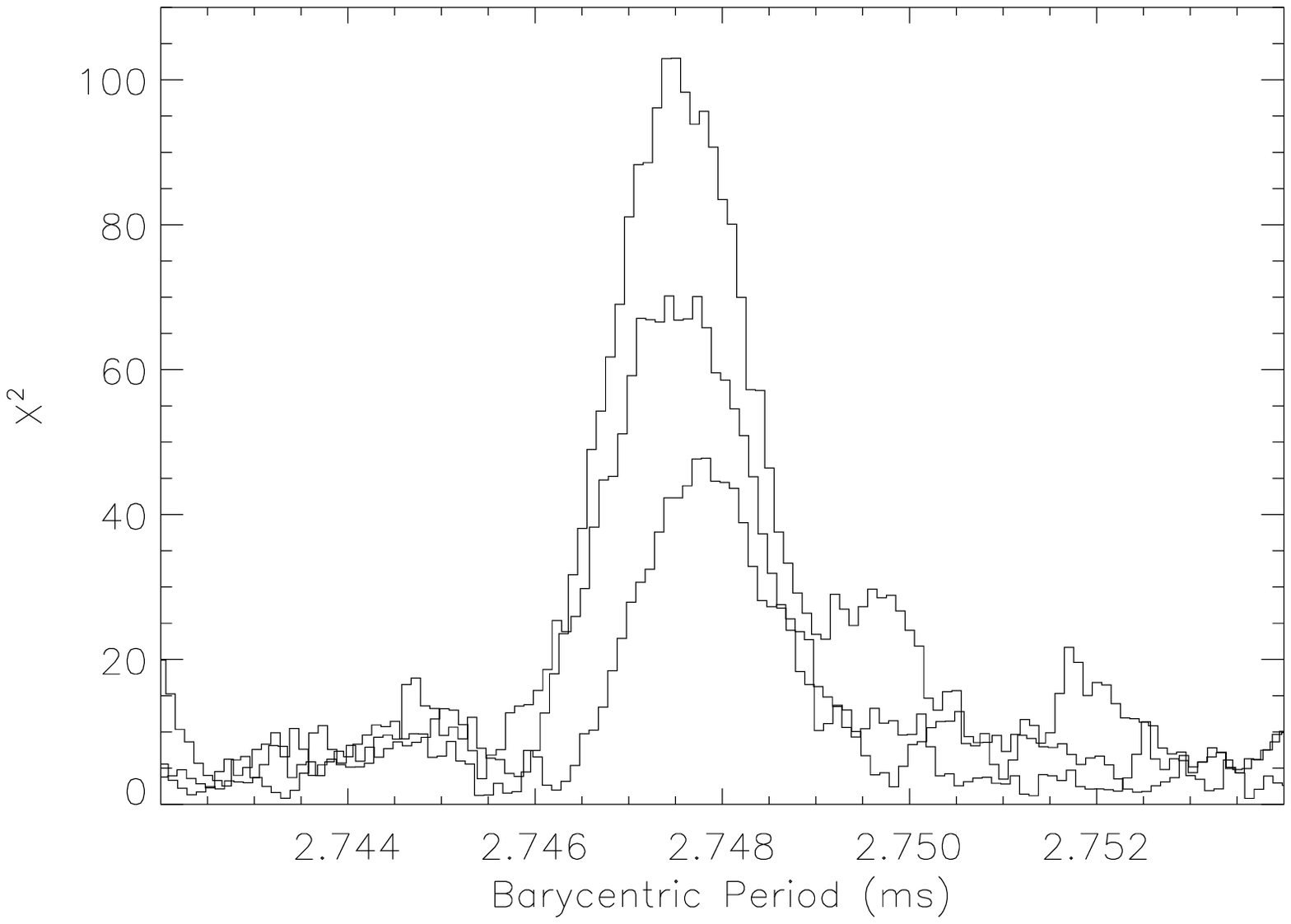,height=5.5in,
width=5.5in}}
\vspace{10pt}
\caption{Results from the $\chi^2$ epoch folding analysis for the
three bursts from 4U 1728-34. The bursts are arranged in time order from bottom
to top (burst 1 at bottom to burst 3 at top). See Table 1 for the measured 
period centroids and uncertainties.}
\label{fig2}
\end{figure*}

\vfill\eject

\begin{figure*}[htb] 
\centerline{\epsfig{file=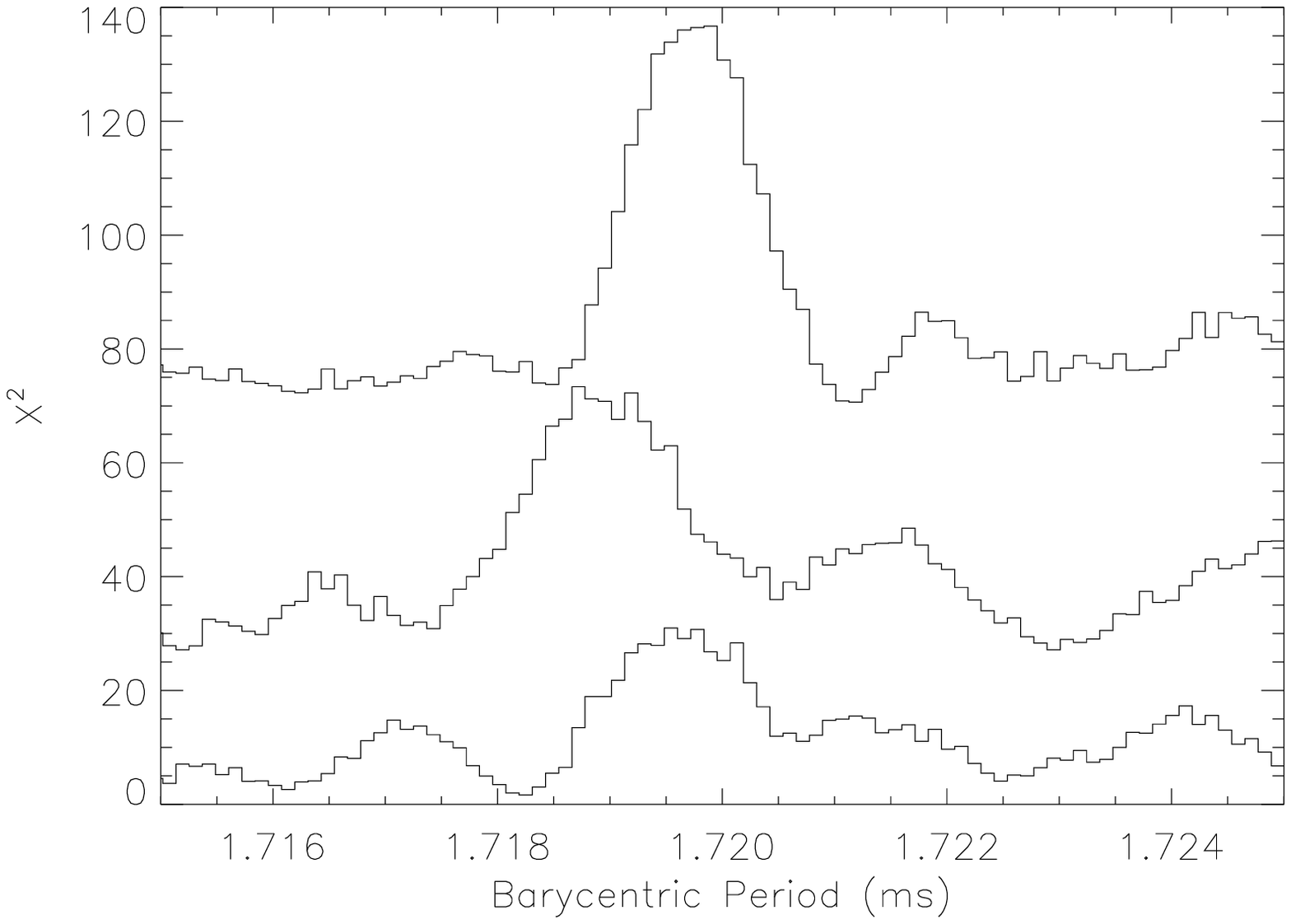,height=5.5in,
width=5.5in}}
\vspace{10pt}
\caption{Results from the $\chi^2$ epoch folding analysis for the
three bursts from 4U 1636-53. The bursts are arranged in time order from bottom
to top (burst 1 at bottom to burst 3 at top). See Table 1 for the measured 
period centroids and uncertainties.}
\label{fig3}
\end{figure*}

\vfill\eject

\begin{figure*}[htb] 
\centerline{\epsfig{file=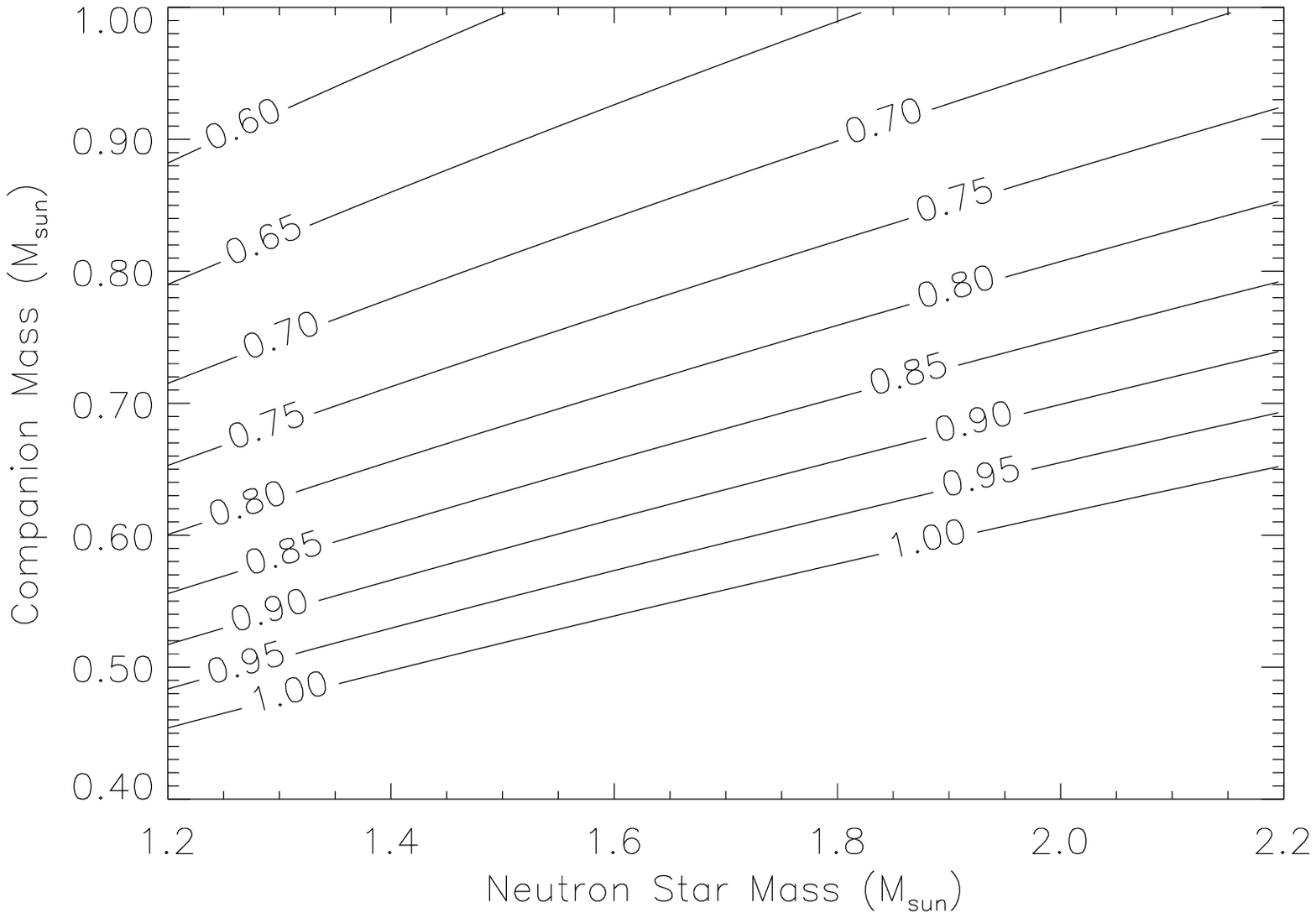,height=5.5in,
width=5.5in}}
\vspace{10pt}
\caption{Contours of constant $\sin i$ for a binary system with 
orbital period 3.8 hr and $v \sin i / c = 4.25 \times 10^{-4}$ as suggested
by the observed period offsets in bursts from 4U1636-53. This result should
not yet be taken as a constraint on the system masses in 4U 1636-53, rather it
only suggests that the orbital motion is a plausible explanation for the
observed period shifts.}
\label{fig4}
\end{figure*}


\begin{references}

\reference{BCUC98} Bildsten, L., Cumming, A., Ushomirsky, G. \& Cutler, C. 1998,
to appear in Proceedings of ``A Half Century of Stellar Pulsation
Interpretations", ASP Conference Ser. (astro-ph/9712358)

\reference{P84}Patterson, J. 1984, ApJS, 54, 443

\reference{PvL81}Pedersen, H., van Paradijs, J., \& Lewin, W. H. G. 1981,
Nature, 294, 725

\reference{SM88}Smale, A. P., \& Mukai, K. 1988, MNRAS, 231, 663

\reference{SMB} Smith, D., Morgan, E. H. \& Bradt, H. V. 1997, ApJ, 479, L137

\reference{SZS}Strohmayer, T. E., Zhang, W. \& Swank, J. H. 1997, ApJ, 487, L77

\reference{S2} Strohmayer, T. E. 1992, ApJ, 388, 138

\reference{Stroh96} Strohmayer, T. E., Zhang, W., Swank, J. H., Smale, A. P.,
Titarchuk, L., Day, C. \& Lee, U. 1996, ApJ, 469, L9

\reference{Stroh97} Strohmayer, T. E., Jahoda, K., Giles, A. B. \& Lee, U. 1997,
ApJ, 486, 355

\reference{vP90} van Paradijs, J., {\it et al.} 1990, A\&A, 234, 181

\reference{Z96}Zhang, W., Lapidus, I., Swank, J. H., White, N. E. \&
Titarchuk, L. 1996, IAUC 6541

\reference{Z98}Zhang, W., Jahoda, K., Kelley, R. L., Strohmayer, T. E., Swank,
J. H. \& Zhang, S. N.

\end{references}
\end{document}